\documentclass{aa}  
\usepackage{natbib}
\usepackage{graphicx}
\usepackage{txfonts}
\usepackage[normalem]{ulem}
\usepackage{xcolor}
\usepackage{placeins} 
\usepackage{caption}
\usepackage{placeins} 
\usepackage{wrapfig}
\usepackage{float}

\bibpunct{(}{)}{;}{a}{}{,} 
\definecolor{links}{rgb}{0, 0, 255}
\usepackage[colorlinks=true,allcolors=links]{hyperref}
\usepackage{txfonts}

\newcommand{\mpchi}{\,h^{-1}{\rm {Mpc}}}

\newcommand{\kms}{\mathrm{km\,s^{-1}}}
\newcommand{\msun}{M_{\sun}}

\newcommand{\hi}{H{~\sc i}}

\begin{document} 

\title{Disparate effects of circumgalactic medium angular momentum in IllustrisTNG and SIMBA}
\titlerunning{Angular Momentum of Circumgalactic Medium}

\author{
Kexin Liu\inst{\ref{inst1},\ref{inst2}}
\and Hong Guo\inst{\ref{inst1}}\fnmsep\thanks{Corresponding Author}
\and Sen Wang\inst{\ref{inst3}} 
\and Dandan Xu\inst{\ref{inst3}}
\and Shengdong Lu\inst{\ref{inst4}}
\and Weiguang Cui\inst{\ref{inst5},\ref{inst6},\ref{inst7}}
\and Romeel Dav\'e\inst{\ref{inst7},\ref{inst8},\ref{inst9}}
}

\institute{
Shanghai Astronomical Observatory, Chinese Academy of Sciences, Shanghai 200030, China.\label{inst1} \email{guohong@shao.ac.cn}
\and University of Chinese Academy of Sciences, Beijing 100049, China.\label{inst2} \email{liukexin@shao.ac.cn}
\and Department of Astronomy, Tsinghua University, Beijing 100084, China.\label{inst3}
\and Institute for Computational Cosmology, Department of Physics, University of Durham, South Road, Durham, DH1 3LE, UK.\label{inst4}
\and Departamento de Física Teórica, M-8, Universidad Autónoma de Madrid, Cantoblanco E-28049, Madrid, Spain.\label{inst5}
\and Centro de Investigación Avanzada en Física Fundamental (CIAFF), Universidad Autónoma de Madrid, Cantoblanco, E-28049 Madrid, Spain.\label{inst6}
\and Institute for Astronomy, University of Edinburgh, Royal Observatory, Edinburgh EH9 3HJ, UK.\label{inst7}
\and University of the Western Cape, Bellville, Cape Town 7535, South Africa.\label{inst8}
\and South African Astronomical Observatories, Observatory, Cape Town 7925, South Africa.\label{inst9}
}

\abstract
  {In this study, we examine the role of the circumgalactic medium (CGM) angular momentum ($j_{\rm CGM}$) on star formation in galaxies, whose influence is currently not well understood. The analysis utilises central galaxies from two hydrodynamical simulations, SIMBA and IllustrisTNG. We observe a substantial divergence in how star formation rates correlate with CGM angular momentum between the two simulations. Specifically, quenched galaxies in IllustrisTNG show higher $j_{\rm CGM}$ than their star-forming counterparts with similar stellar masses, while the reverse is true in SIMBA. This difference is attributed to the distinct active galactic nucleus (AGN) feedback mechanisms active in each simulation. Moreover, both simulations demonstrate similar correlations between $j_{\rm CGM}$ and environmental angular momentum ($j_{\rm Env}$) in star-forming galaxies, but these correlations change notably when kinetic AGN feedback is present. In IllustrisTNG, quenched galaxies consistently show higher $j_{\rm CGM}$ compared to their star-forming counterparts with the same $j_{\rm Env}$, a trend not seen in SIMBA. Examining different AGN feedback models in SIMBA, we further confirm that AGN feedback significantly influences the CGM gas distribution, although the relationship between the cold gas fraction and the star formation rate (SFR) remains largely stable across different feedback scenarios.}

   
   \maketitle
%

\section{Introduction}

The structure and star formation activity of a galaxy are closely linked to its kinematic properties. It has been proposed that disc galaxies are formed from gas accretion and acquire angular momentum from tidal torques \citep[see e.g.,][]{Fall1980,Barnes1987,Mo1998,Bullock2001,Keres2005}. On the other hand, elliptical galaxies have a factor of five less angular momentum than disc galaxies of similar masses \citep[e.g.,][]{Fall1983,Romanowsky2012,Fall2013}. 
	
Thanks to large-scale integral field spectroscopy (IFS) surveys, galaxies can be further classified by their dynamical features rather than by simple morphology or colour \citep{Driver2007}. The SAURON project \citep{Bacon2001}, which is based on the IFS, complements the traditional classification of early-type galaxies (ETGs) by including information about galaxy kinematics. The ETGs are divided into fast and slow rotators in the anisotropy diagram between the ratio of ordered and random motion ($V/\sigma$) and ellipticity ($\rm \varepsilon$) \citep[see e.g.,][]{Cappellari2007,Emsellem2007}. 
\cite{van2018} further found a strong correlation between the stellar population age and the location in the ($V/\sigma$, $\rm \varepsilon$) diagram, where fast rotators are younger and slow rotators are relatively old. This suggests that the star formation history is correlated with galaxy stellar kinematics. 

The circumgalactic medium (CGM) is believed to be a source of fuel for star formation and is thus expected to be linked to star formation activity. Using the IllustrisTNG hydrodynamical simulation \citep[hereafter TNG;][]{Pillepich2018,Nelson2019}, \cite{Lu2022} found that quenched early-type galaxies typically have higher specific angular momenta than their star-forming disc counterparts with similar stellar masses. This high angular momentum of the CGM could impede the efficient radial gas inflow that is necessary to maintain star formation in the centre and is thus seen as an additional factor in keeping galaxies quenched, a process known as angular momentum quenching \citep{Peng2020}. The high angular momenta of quenched galaxies in TNG are probably related to the more frequent prograde merger events (relative to the spin of the galaxy itself, as the angular momentum of the CGM gas is inherited from the large-scale halo environment \citep{Lu2022,Wang2022}.

However, the angular momentum and gas distributions of galaxies in hydrodynamical simulations can be easily altered by the baryon physics employed. Negative feedback from the active galactic nucleus (AGN) is one of the most effective ways to quench galaxies in these simulations. \cite{Ma2022} compared the hydrodynamical simulations of TNG and SIMBA \citep{Dave2019} and found that the kinetic AGN feedback in TNG only moves cold gas from the inner discs to the outer regions, thus reducing the star formation activities in the inner regions. This leads to TNG overestimating the overall cold-gas reservoir in quenched galaxies compared to observations. Therefore, angular momentum quenching is necessary in TNG to prevent the CGM gas of these quenched galaxies from falling into the centre. On the other hand, the bipolar injection of kinetic AGN feedback in SIMBA only slightly reduces the cold gas densities in the inner regions, but significantly heats the CGM gas, which brings SIMBA into better agreement with the observed cold-gas masses for quenched galaxies. Without a consistent supply of cold gas in the CGM, quenched galaxies will remain with low star formation rates, so it does not seem necessary to have the additional input of angular momentum quenching. 

It is perplexing to consider the impact of angular momentum on star formation in galaxies when comparing the two simulations. To gain a better understanding of the effect of angular momentum in different hydrodynamical simulations, this paper will explore the relationship between star formation rate (SFR) and CGM angular momentum in TNG and SIMBA. In addition, we will examine the influence of various morphologies on the angular momentum distribution. The paper is organised as follows. Section~\ref{sec:data} provides an overview of the galaxy samples in the two simulations. The main findings are summarised in Section~\ref{sec:results}. Conclusions and discussions are described in Sections~\ref{sec:conclusion}.

\section{Simulation and data}\label{sec:data}
	
\subsection{The IllustrisTNG simulation}
The IllustrisTNG project consists of a set of magnetohydrodynamical simulations of different cosmological volumes \citep{Marinacci2018,Naiman2018,Nelson2018,Nelson2019,Pillepich2018,Springel2018}, run with the Arepo code \citep{Springel2010}. In this paper, we focus on the TNG100-1 simulation with a box size of $75\mpchi$ on a side, and the cosmological parameters are $\Omega_{\rm m}=0.3089$, $\Omega_{\lambda}=0.6911$, $\Omega_{\rm b}=0.0486$, and $h=0.6774$. The dark matter mass resolution is $m_{\rm DM}=7.5 \times 10^{6}\msun$ and the average gas cell mass is around $m_{\rm gas}=1.4 \times 10^{6}\msun$.

The Illustris simulation \citep{Vogelsberger2014} has been improved by the TNG model with the introduction of two AGN feedback modes: thermal and kinetic. These two modes are exclusive and are determined by comparing the Eddington ratio, $\rm \lambda_{Edd}$, with the critical value, $\rm \chi$, which is calculated using equation $\rm \chi=\rm{min} \left[ 0.002(M_{\rm BH}/10^{8}\msun)^2,0.1 \right]$. When the Eddington ratio is greater than or equal to the critical value, the thermal mode is activated, which injects energy isotropically into the environment and heats the surrounding gas. On the other hand, when the Eddington ratio is less than the critical value, the AGN feedback switches to the kinetic mode, which produces black hole-driven winds and transfers energy to the neighbouring gas in the form of random momentum \citep{Weinberger2017}. This kinetic mode has enabled the stellar mass fraction in massive haloes to be more in line with observations, without the overheating of gas as in Illustris.

\subsection{The SIMBA simulation}\label{subsec:simba}
SIMBA simulations \citep{Dave2019} are carried out using the GIZMO code \citep{Hopkins2015}, which is a solver for cosmological gravity and hydrodynamics. This paper adopts the flagship run of SIMBA (m100n1024), which has a box size of $100\mpchi$ on a side and cosmological parameters that are consistent with TNG, including $\Omega_{\rm m}=0.3$, $\Omega_{\lambda}=0.7$, $\Omega_{\rm b}=0.048$, and $h=0.68$. The resulting dark matter and gas cell mass resolutions are $9.6 \times10^{7}\msun$ and $1.82\times 10^{7}\msun$, respectively. Galaxies in SIMBA are identified by a 6D friends-of-friends (FOF) finder with a linking length of 0.0056 times the mean inter-particle distance, while haloes are identified using a 3D FoF finder with a linking length parameter of 0.2.

SIMBA assumes the H$_2$-based star formation law \citep{Schmidt1959}, where the H$_2$ fraction is determined by local metallicity and gas column density \citep{Krumholz2011}. The star formation rate (SFR) is then calculated using the equation: ${\rm SFR}= \varepsilon_{\ast} \rho_{\rm H_2}/t_{\rm dyn}$, where $\varepsilon_{\ast}=0.02$ \citep{Kennicutt1998} and the dynamical time $t_{\rm dyn}=1/ \sqrt{G \rho}$. The \hi\ gas fraction is calculated based on the prescription of \cite{Rahmati2013}. The total neutral gas can then be calculated on the fly. Each gas cell in haloes is assigned to the galaxy with the highest value of $M_{\rm baryon}/R^{2}$, where $M_{\rm baryon}$ is the total baryonic mass of the galaxy and $R$ is the distance from the gas cell to the centre of the galaxy. SIMBA incorporates the star formation-driven feedback from MUFASA and a torque-limited accretion model \citep{Hopkins2011,Angles2017} to accrete cold gas ($T<10^{5}$~K), which does not require self-regulation for black hole growth \citep{Angles2015}. For hot gas ($T>10^{5}$~K), Bondi accretion is used \citep{Bondi1952}, which is the same as in TNG.
	
In SIMBA, two main kinetic feedback models are present, along with an X-ray feedback model. When the Eddington ratio is high ($\lambda_{\rm Edd} > 0.2$), a radiative or wind mode is activated, which propels the gas away from the black hole at speeds of up to $1000\,\kms$. When the Eddington ratio is low ($\lambda_{\rm Edd} < 0.2$) and the black hole mass is greater than $10^{7.5}\msun$, a jet mode is triggered, which can reach velocities of up to $8000\,\kms$ when $\lambda_{\rm Edd} < 0.02$. Both the jet and wind modes are bipolar and purely kinetic, and a small fraction of the energy of the jet mode will increase the virial temperature of the halo. When the full jet condition ($\lambda_{\rm Edd} < 0.02$) and $ f_{\rm H_{2}} < 0.2$ is met, the X-ray heating mode is activated. This mode has the least effect on galaxies \citep{Dave2019}, and is usually insignificant.
	
The SIMBA project has a set of simulations with a box size of $50 \mpchi$ on a side (simulation name m50n512), containing $2\times512^3$ particles (i.e. the same resolution and input physics as the flagship run m100n1024), but with a volume eight times smaller. In addition to the full m50n512 run, there are catalogues that have different AGN feedback mechanisms turned off: `No-Xray', `No-jet', and `No-AGN', which correspond to X-ray feedback off, both X-ray and jet models off, and all AGN feedback models off, respectively. We use both m100n1024 and m50n512 runs to investigate the effect of CGM angular momentum with different feedback conditions.
	
\subsection{Sample selection} \label{sec:method}
	
\begin{figure}
\centering
\includegraphics[width=\columnwidth]{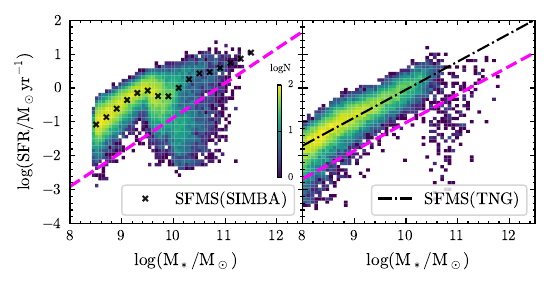}
\caption{Distribution of central galaxies at $z=0$ as a function of SFR and $\rm M_\ast$, for SIMBA and TNG in 100Mpc/h, from left to right, respectively. The coloured background is the logarithmic number of central galaxies . The black dot-dashed line is the SFMS of TNG100, following the definition in \cite{Ma2022}: $\log({\rm SFR}/{\rm yr}^{-1}\msun )=0.83\log \rm M_\ast-8.32$, and black cross represents SFMS of star-forming central galaxies in SIMBA100, where star-forming galaxies are simply selected with $\log({\rm sSFR}/{\rm Gyr}^{-1}\msun) \geq -1.8$. Magenta lines separate the star-forming and quenched galaxies in both panels.}\label{fig:sfms}
\end{figure}
 
In this paper, we focus exclusively on the central galaxies in both TNG and SIMBA. The SFR in TNG is determined from the value of `SubhaloSFR', that is, the total SFR of all gas cells in a given subhalo. In SIMBA, the SFR is defined as the total SFR in gas particles belonging to each galaxy. Figure~\ref{fig:sfms} displays the central galaxy distributions in the ${\rm SFR}$--$M_\ast$ plane, with the colour indicating the number density of central galaxies. We use the definitions of the star formation main sequence (SFMS) ${\rm SFR_{MS}}(M_\ast)$ in SIMBA (left panel) and TNG (right panel) from \cite{Ma2022}, which are shown as the crosses and the dot-dashed line, respectively. Since the SFMS of SIMBA is curved for $M_{\ast} > 10^{9.5}\msun$, we only consider galaxies with stellar masses ranging from $10^{9.5}\msun$ to $10^{11.2}\msun$ in both simulations at $z=0$ to ensure that the galaxies in our sample are fully resolved.

Finally, we select a total of 6925 and 14479 central galaxies in the mass range of $\rm 9.5 \leq \log(M_\ast/\msun) \leq 11.2$ from SIMBA and TNG, respectively. For a fair comparison between the two simulations, we focus on the SFR offsets from the SFMS, $\Delta\log{\rm SFR}$, defined as,
\begin{equation}
  \Delta\log{\rm SFR}=\log{\rm SFR}-\log{\rm SFR}_{\rm MS}.
\end{equation}
Following \cite{Ma2022}, we separate the star-forming and quenched galaxies with the cutoff SFR of $\Delta\log{\rm SFR}=-1$, that is, 1~dex below the SFMS.

\subsection{Kinematic properties}\label{sec:method}

According to \cite{Lu2022}, we consider the region beyond twice the half stellar mass radius of the galaxy ($2R_{\rm hsm}$) and within $100$~kpc to be the CGM domain. Beyond this scale, the connection between CGM cold gas and inner star formation is weak. We focus on the cold phase of the CGM gas, which has a temperature range of $10^{4}{\rm K}<T<2\times10^{4}$~K. In \cite{Wang2022}, it was found that the cold-phase CGM gas has a higher angular momentum and more tangential motion than its hot-phase ($T>10^{5}$~K) counterparts. 

In both TNG and SIMBA simulations, we calculate the specific angular momentum of the cold-phase CGM gas ($10^{4}{\rm K}<T<2\times10^{4}$~K), $\mathbf{j}_{\rm CGM}$, as follows,
\begin{equation} 
\mathbf{j}_{\rm CGM} = \frac{\sum_{i} m_{{\rm gas},i}\mathbf{r}_{{\rm gas},i} \times \mathbf{v}_{{\rm gas},i}}{\sum_{i} m_{{\rm gas},i}},
\end{equation}
where $m_{{\rm gas},i}$ is the mass of a given gas cell $i$, $\mathbf{r}_{{\rm gas},i}$ is the relative distance from the galaxy centre, and $\mathbf{v}_{{\rm gas},i}$ is the velocity. The summation is performed in all cold gas cells with $2R_{\rm hsm}<r_{{\rm gas},i}<100$kpc. The dimensionless spin of CGM ($\lambda_{\rm CGM}$) can be measured as follows \citep{Lu2022},
\begin{equation}
{\lambda}_{\rm CGM} =\frac{\Sigma_i m_{{\rm gas},i}v^2_{{\rm t,gas},i}r_{{\rm gas},i}/GM(\leq r_{{\rm gas},i})}{\Sigma_i m_{{\rm gas},i}}  \label{eq:}
\end{equation}
where $v_{{\rm t,gas},i}$ is the tangential velocity of the $i$-th gas cell and $M(\leq r_{{\rm gas},i})$ is the total mass (including all types of particles) within $r_{{\rm gas},i}$. 

Different from \cite{Lu2022}, we calculate the environmental angular momentum of each central galaxy, $\mathbf{j}_{\rm env}$, by summing the contribution from all particles (including dark matter, star, and gas particles) as,
\begin{equation}
\mathbf{j}_{\rm env} =\frac{\Sigma_{i} M_{{\rm par},i}\mathbf{r}_{i} \times \mathbf{v}_i}{\Sigma_i M_{{\rm par},i}},
\end{equation}
where $M_{{\rm par},i}$ is the mass of an individual particle $i$, $\mathbf{r}_{i}$ is the relative distance from $i$ to the central galaxy, and $\mathbf{v}_i$ is the relative velocity. The summation is performed on all particles with $2R_{\rm hsm}<r_i<100$kpc, where dark matter will dominate the contribution of angular momentum. This will ensure a better quantification of the environmental angular momentum than just using the satellite galaxies as in \cite{Lu2022}. 
	
\section{Results}\label{sec:results}
\subsection{Trend with SFR}
	
\begin{figure}
\centering
\includegraphics[width=0.9\columnwidth]{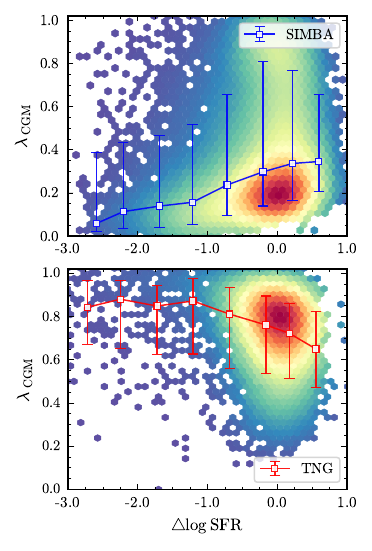}
\caption{Distribution of $\lambda_{\rm CGM}$ as a function of $\Delta\log{\rm SFR}$ in SIMBA (top panel) and TNG (bottom panel). The background is the logarithmic number of galaxies, with colours ranging from light (blue) to dark (red) indicating an increasing number of galaxies. From top to bottom panels, the blue and red lines show the median values and errors estimated from the $16^{th}-84^{th}$ percentile ranges for SIMBA and TNG, respectively.}
\label{fig:spinsfr}
\end{figure}

In Figure~\ref{fig:spinsfr}, we show the distribution of galaxies in the $\lambda_{\rm CGM}$--$\Delta\log{\rm SFR}$ plane for SIMBA (top panel) and TNG (bottom panel). The colour coding indicates the logarithmic number of galaxies. The solid line in each panel indicates the median values with errors estimated from the $16^{\rm th}$--$84^{\rm th}$ percentile ranges. The two simulations show different correlations between $\lambda_{\rm CGM}$ and $\Delta\log{\rm SFR}$. In TNG, galaxies with a low $\Delta\log{\rm SFR}$ have much higher $\lambda_{\rm CGM}$ than their star-forming counterparts. This is in agreement with the findings of \cite{Lu2022}, which suggest that the CGM gas of the quenched galaxies still has a strong rotational motion. It is likely related to the surplus of cold gas in the CGM caused by the relatively weak kinetic AGN feedback in the TNG \citep{Ma2022}. 

On the other hand, $\lambda_{\rm CGM}$ gradually increases with $\Delta\log{\rm SFR}$ in SIMBA and most quenched galaxies only have a small $\lambda_{\rm CGM}$ ($<0.2$), indicating that the CGM gas in quenched galaxies is more likely to have random motion rather than rotational motion. This is also in line with their low cold gas reservoir, as reported in \cite{Appleby2021} and \cite{Ma2022}. The majority of galaxies with high-$\lambda_{\rm CGM}$ are the disc galaxies in the SFMS.

\begin{figure}
\centering
\includegraphics[width=0.9\columnwidth]{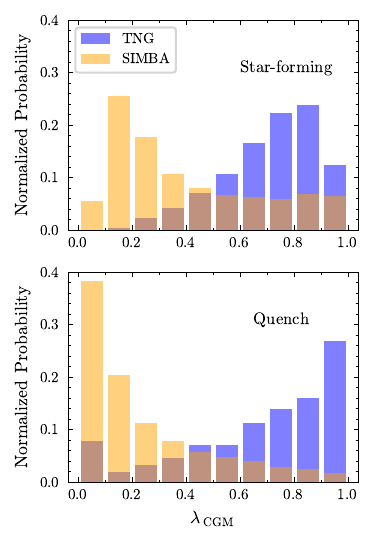}
\caption{Histograms of $\lambda_{\rm CGM}$ for SIMBA and TNG, with the top and bottom panels representing star-forming and quenched galaxies, respectively, where quenched galaxies include those with ${\rm SFR}=0$, and the Y-axis is normalised by the total number of central galaxies in each simulations. Blue and orange bins represent normalised probability of TNG and SIMBA, respectively, and the brown is cross area.}
\label{fig:sfr0}
\end{figure}
We note that the comparisons in Figure~\ref{fig:spinsfr} do not include galaxies without SFR measurements (i.e. ${\rm SFR}=0$) in both simulations. The fraction of galaxies with ${\rm SFR}=0$ is 3.5\% and 9\% for SIMBA and TNG, respectively. These galaxies are caused by the stochastic star formation algorithms and the limited simulation resolution. So they can be effectively considered as fully quenched. To be comprehensive, Figure~\ref{fig:sfr0} displays the probability distributions of $\lambda_{\rm CGM}$ for all quenched galaxies (including galaxies with ${\rm SFR}=0$) and star-forming galaxies in SIMBA (orange) and TNG (blue). Both the star-forming and quenched galaxies show very different distributions in $\lambda_{\rm CGM}$ between SIMBA and TNG. Star-forming galaxies in SIMBA have significantly lower $\lambda_{\rm CGM}$ compared to TNG. It is evident that the $\lambda_{\rm CGM}$ distributions for all quenched galaxies still follow the trend for quenched galaxies shown in Figure~\ref{fig:spinsfr}, whether in TNG or SIMBA. Although galaxies with ${\rm SFR}=0$ in TNG have slightly lower $\lambda_{\rm CGM}$ values (with an average of around 0.72) than the median values ($\sim 0.8$) shown in Figure~\ref{fig:spinsfr}, they are still comparable to the $\lambda_{\rm CGM}$ values of star-forming galaxies (with an average of 0.76), as illustrated in Figure~\ref{fig:sfr0}. The situation is similar in SIMBA. Those with ${\rm SFR}=0$ in SIMBA have an average $\lambda_{\rm CGM}$ of $0.11$, while those star-forming galaxies have an average value of $0.28$.

\begin{figure*}
\centering
\includegraphics[width=\textwidth]{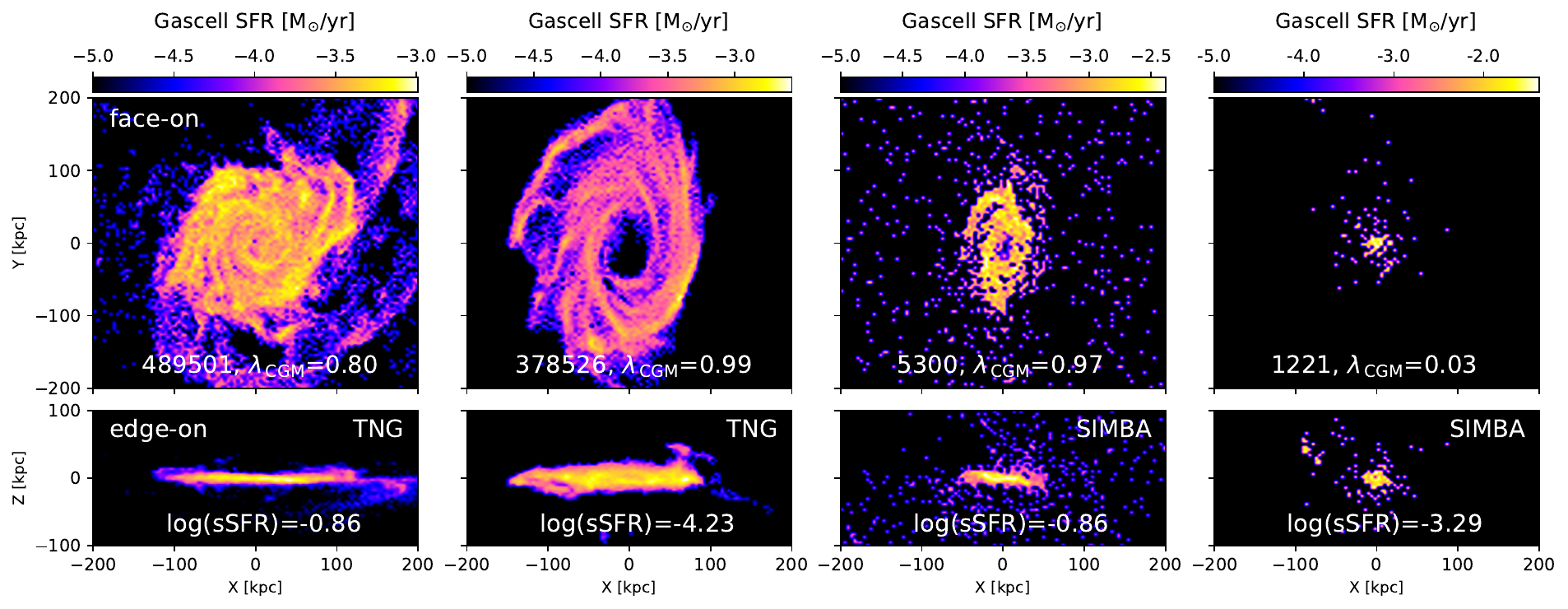}
\caption{Spatial distribution of cold gas in four examples of TNG and SIMBA, respectively. Stellar masses of all four galaxies are in the range of $ 10^{10.2} - 10^{10.8}\msun$. Top panels are shown in face-on projection, along with the direction of galaxy gas angular momentum, and bottom is in edge-on projection. The gas distribution is colour-coded by its star-formation rate, and $\lambda_{\rm CGM}$ and $\log$ sSFR $(\rm M_{\odot}/Gyr)$ are labelled in each panel.}
\label{fig:examples}
\end{figure*}
Figure~\ref{fig:examples} presents examples of cold gas distributions in star-forming and quenched galaxies in TNG (two left panels) and SIMBA (two right panels). The top and bottom panels display face-on and edge-on projections, respectively. The colour scales indicate the SFRs of the gas cells, as labelled. The total specific SFR of each galaxy (${\rm sSFR}={\rm SFR}/M_\ast$) is also shown in the panel. Notably, both the star-forming and quenched galaxies in TNG have large cold-gas discs in the CGM, which is in agreement with their high angular momenta, as seen in Figure~\ref{fig:spinsfr}. However, the quenched galaxies have an obvious cavity of cold gas in the central region (roughly the size of the stellar disc). As demonstrated in \cite{Ma2022}, galaxies in TNG are quenched by the ejection of cold gas from inner discs to the outer regions, which is caused by the kinetic AGN feedback mode. This naturally creates a hole in the gas distribution and halts star formation in the central stellar disc. 

In contrast to star-forming galaxies, quenched galaxies in SIMBA have very few cold gas cells, and their distribution is more spherical than disc-like. Additionally, there is no visible cavity of cold gas in the core. The cold gas density in the central parts of star-forming and quenched galaxies in SIMBA are quite similar \citep[see Fig.~5 of][]{Ma2022}. However, the quenched galaxies have much less cold gas in the CGM, which reduces the amount of cold gas that can reach the core, and, as a result, the overall SFR decreases. 

It is evident from the comparisons that galaxies in TNG and SIMBA, despite having similar stellar masses and star formation rates, have distinct CGM gas kinematics that are linked to the total cold gas supply and their AGN feedback processes.

\subsection{Cold gas reservoir}
\begin{figure*}
\centering
\includegraphics[width=\textwidth]{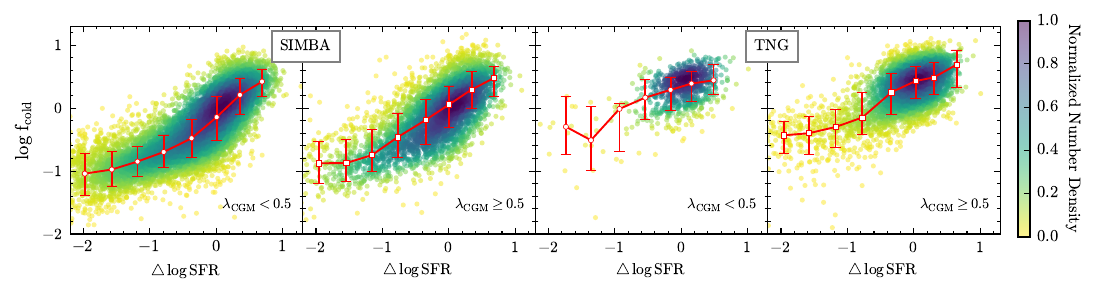}
\caption{Relations between $\log \rm f_{\rm cold}$ and $\Delta\log{\rm SFR}$, for TNG and SIMBA, from left to right, respectively. The coloured background is the probability density distribution, as described in Figure~\ref{fig:spinsfr}. The solid red line is plotted by median values with errors estimated from the $16^{th}-84^{th}$ percentile ranges, and is simply distinguished by high and low spin. }\label{fig:fgassfr}
\end{figure*}

To be clear, we compute the cold gas mass in the CGM, $M_{\rm gas}$, by summing the cold gas cells within the range of $2R_{\rm hsm}$ to $100$~kpc. Additionally, we define $f_{\rm cold}\equiv M_{\rm gas}/M_\ast$ as the ratio of $M_{\rm gas}$ to the total stellar mass of galaxies ($M_\ast$). The relationship between $f_{\rm cold}$ and $\Delta\log{\rm SFR}$ is analysed in Figure~\ref{fig:fgassfr}, with SIMBA displayed in the two left panels and TNG in the two right panels. The colour scales, as indicated by the colour bar on the right, represent the normalised galaxy number density in each simulation. We categorise the galaxy distributions into subsamples of $\lambda_{\rm CGM}<0.5$ and $\lambda_{\rm CGM}>0.5$, as labelled in each panel. It is apparent that in SIMBA, angular momentum scarcely affects the correlation between SFR and the CGM gas mass. In most quenched galaxies, the CGM shows low $f_{\rm cold}$ and low angular momentum. In TNG, the decline in $\Delta\log{\rm SFR}$ with decreasing $f_{\rm cold}$ reflects the trend seen in SIMBA. For quenched galaxies, the association between $\Delta\log{\rm SFR}$ and $f_{\rm cold}$ is considerably weaker because the cold gas in CGM is not actively involved in star formation. However, there are very few quenched galaxies with $\lambda_{\rm CGM}<0.5$, which is in stark contrast to the results in SIMBA. The physical mechanisms behind these differences will be examined in Section~\ref{subsec:feedback}.

\subsection{Environmental effect}
\begin{figure}
		\centering
        \includegraphics[width=0.8\columnwidth]{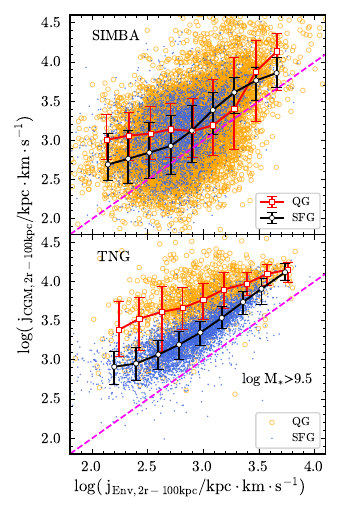}
    
		\caption{Relations between $\rm j_{CGM}$ and $\rm j_{Env}$ within 2$R_{\rm hsm}$ to 100 kpc, for SIMBA and TNG, from top to bottom, respectively. The black line is the relation of SFGs and red line is the relation of QGs in each panel. To show the trend more clearly, we present 1:1 baselines, indicated by the magenta dotted lines. For SFGs and QGs, each sample is plotted with a blue dot and an orange circle, respectively. Each line shows the median values, with errors estimated from the $16^{th}-84^{th}$ percentile ranges.}
		\label{fig:jenv}
\end{figure}

According to classical galaxy formation theory, galaxies gain angular momentum by accreting gas from their host halos as they evolve, influenced by the torque field \citep{Peebles1969,Mo1998}, consistent with findings in \cite{Lu2022}. Figure~\ref{fig:jenv} illustrates the relationship between the angular momentum of the CGM ($j_{\rm CGM}$) and that of the environment ($j_{\rm Env}$) within the range of $\rm 2R_{hsm}$--$100\,{\rm kpc}$ for galaxies of $\log(M_\ast/\msun)>9.5$. The top and bottom panels reveal the results for SIMBA and TNG simulations, respectively, distinguishing star-forming galaxies (SFGs, shown by blue points) from quenched galaxies (QGs, depicted by orange circles). The median values of $j_{\rm CGM}$ for the two populations in different $j_{\rm Env}$ bins are shown with solid lines. It is apparent that the CGM angular momentum, $j_{\rm CGM}$, is influenced by the environmental angular momentum, $j_{\rm Env}$, for both SFGs and QGs in both simulations. This indicates that the galaxy's angular momentum is largely inherited from its surrounding environment. But the correlation between $j_{\rm CGM}$ and $j_{\rm Env}$ decreases at lower values of $j_{\rm Env}$. 

Interestingly, the SFGs in two different simulations demonstrate a similar relationship between $j_{\rm CGM}$ and $j_{\rm Env}$. The angular momentum of the CGM is influenced by both the inner stellar disc and the outer halo environment. When $j_{\rm Env}$ is large, the environment primarily contributes to $j_{\rm CGM}$. Consequently, the SFGs and QGs in both simulations converge for $\log(j_{\rm Env}/{\rm kpc}\,{\rm km\,s^{-1}})>3.5$, with a slope of $j_{\rm CGM}/j_{\rm Env}$ near unity. Conversely, at low $j_{\rm Env}$ values, the inner disc's influence becomes significant, making the $j_{\rm CGM}/j_{\rm Env}$ slope approach zero. This accounts for the behaviour of the SFGs in the simulations, which are likely to be significantly influenced by powerful AGN feedback. We note that $j_{\rm CGM}$ and $j_{\rm Env}$ cover the same range of CGM ($2R_{\rm hsm}$--$100$~ kpc) and, in principle, are closely connected. It would be intriguing to compare the relation between $j_{\rm CGM}$ and $j_{\rm gal}$, where $j_{\rm gal}$ is the angular momentum calculated using all particles within $2R_{\rm hsm}$ of each galaxy. We find that $j_{\rm Env}$ and $j_{\rm gal}$ are closely correlated with each other, and the relations between $j_{\rm CGM}$ and $j_{\rm gal}$ still largely follow the trends between $j_{\rm CGM}$ and $j_{\rm Env}$.

The primary difference between the two simulations is that QGs in the TNG simulation consistently exhibit higher $j_{\rm CGM}$ compared to their star-forming counterparts at a specific $j_{\rm Env}$, a trend also observed in \cite{Lu2022}. However, the median values of $j_{\rm CGM}$ for QGs in the SIMBA simulation are roughly the same as those of SFGs. The distribution of $j_{\rm CGM}$ for QGs is significantly broader. This distinct behaviour of QGs in the two simulations is most likely caused by the predominant impact of the AGN feedback mechanisms.
	
\subsection{Effect of AGN feedback}\label{subsec:feedback}
\begin{figure*}
		\centering
		\includegraphics[width=0.9\textwidth]{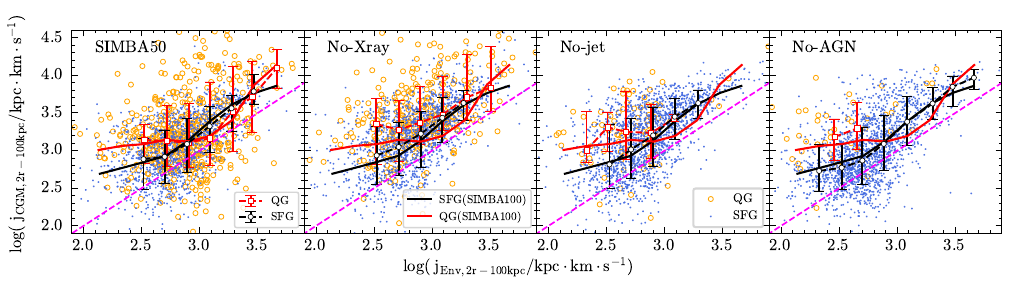}
		\caption{Same as Fig.~\ref{fig:jenv}, but for the fiducial SIMBA50 run, No-Xray run, No-jet run, and No-AGN run, respectively. The black and red dashed lines show the relations of SFGs and QGs in SIMBA50, respectively. To clearly illustrate the effect of feedback, we replicate in each panel the medians for SFGs and QGs in SIMBA100 with black and red solid lines, respectively.}
		\label{fig:jenv_simba50}
\end{figure*} 
The SIMBA simulation set comprises three variations of the implementation of the AGN feedback model by disabling different AGN feedback modes, as described in Section~\ref{subsec:simba}. To examine the influence of AGN feedback on the CGM angular momentum, we compare the fiducial model with the different variations. Figure~\ref{fig:jenv_simba50} illustrates the relationship between $j_{\rm CGM}$ and $j_{\rm Env}$ for these model variations. The models displayed from left to right are the fiducial model, `No-Xray', `No-jet', and `No-AGN' runs. These simulations use a smaller volume of $50^3\,h^{-3}{\rm Mpc}^3$, resulting in fewer galaxies compared to the fiducial run in Figure~\ref{fig:jenv}. We show the median values of $j_{\rm CGM}$ in different $j_{\rm Env}$ bins for SFGs and QGs as the black and red dashed lines, respectively. For fair comparisons, we also replicate in each panel the median values of $j_{\rm CGM}$ from SIMBA100 in Figure~\ref{fig:jenv} as solid lines. The median $j_{\rm CGM}$ for SFGs in the SIMBA50 variation models closely follows the fiducial one in SIMBA100, as expected. But there are slightly larger differences for QGs. The scatters of $j_{\rm CGM}$ in QGs are significantly increased when more AGN feedback mechanisms are included, confirming the results shown in Figure~\ref{fig:jenv}.

AGN feedback serves as the principal mechanism to quench galaxies in SIMBA. The `No-AGN' run, which disables all AGN feedback modes, shows only a small number of QGs. It is clear that the $j_{\rm CGM}$ distribution for QGs matches that of SFGs when jet-mode feedback is disabled in both the `No-jet' and `No-AGN' runs. Kinetic jet feedback leads to a notable increase in the scatter of $j_{\rm CGM}$ for QGs. As outlined in \cite{Dave2019} \citep[and further discussed in][]{Ma2022}, the bipolar kinetic feedback in the low-accretion mode ejects gas particles from within the black hole accretion kernel at an outflow velocity of up to $8000\,\kms$. The collimated nature of the outflow restricts its extent to approximately $1\,{\rm kpc}$. As a result, the cold gas distribution in the centres of the galaxies is not severely affected by the jet-mode feedback, as also shown in Fig.~5 of \cite{Ma2022}. However, jet energy is effectively distributed spherically within the CGM gas due to hydrodynamic interactions between the outflow and the CGM gas, thus modifying the distribution $j_{\rm CGM}$ and significantly reducing the cold gas content in the CGM. 

 TNG has significantly different AGN feedback mechanisms. During its low-accretion kinetic feedback phase, gas cells around a black hole are expelled with momentum in random directions. However, the mean momentum is preserved over numerous energy injection events. Despite the isotropic nature of kinetic energy release in TNG, bipolar gas outflows are still present. As discussed in \cite{dylan2019}, the collimated outflow in TNG is largely caused by a `path of least resistance' effect. Kinetic feedback-driven outflows encounter enhanced resistance on the planes of galactic discs and are reshaped into high-velocity collimated outflows perpendicular to the discs. Meanwhile, outflows along the plane of a disc will expel gas from the centre of the galaxy, which explains the void observed in the CGM gas distribution in Figure~\ref{fig:examples}, as well as the extensive cold gas disc in the QG.
 
The influence of kinetic feedback on the CGM gas distribution depends on both the feedback mechanism and the strength of the released energy. Due to the mild energy release, the outflow velocity in TNG can only reach a maximum value around $3000\,\kms$ in very massive galaxies \citep{dylan2019}, which is significantly smaller than the typical outflow velocity (with a maximum value of $8000\,\kms$) in SIMBA \citep{Dave2019}. As reported in \cite{Ma2022}, kinetic AGN feedback causes the cold gas in TNG to migrate from the inner discs to the outer regions while maintaining the overall cold gas mass.  When the cold gas is redistributed in the outer region, due to momentum conservation, the angular momentum of the CGM will increase accordingly, as shown in the right panel of Figure~\ref{fig:jenv}.

\begin{figure*}
\centering
\includegraphics[width=0.9\textwidth]{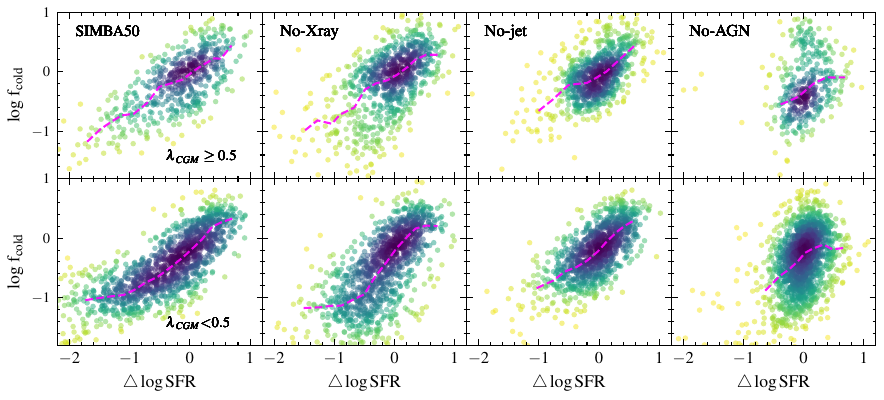}
\caption{Relations between $\Delta\log{\rm SFR}$ and $f_{\rm cold}$, for central galaxies with $\lambda_{\rm CGM}<0.5$ and $\lambda_{\rm CGM} \geq 0.5$ in SIMBA50. Here we show relations of fiducial SIMBA50 run, No-Xray run, No-jet run, and No-AGN run, from left to right, respectively. The background is a probability density distribution of central galaxies, as described in Figure~\ref{fig:spinsfr}. Magenta lines show the change of medians of $f_{\rm cold}$ with the change of $\Delta\log{\rm SFR}$.}
\label{fig:fgas_simba50}
\end{figure*}
To gain a deeper understanding of how CGM angular momentum and AGN feedback influence the gas distribution, Figure~\ref{fig:fgas_simba50} displays the galaxy distributions within the $\log f_{\rm cold}$--$\Delta\log{\rm SFR}$ plane for the various AGN feedback models of SIMBA50. The distributions for galaxies with $\lambda_{\rm CGM}\geq0.5$ and $\lambda_{\rm CGM}<0.5$ are illustrated in the top and bottom panels, respectively. We used the same galaxy mass range and colour scales as in Figure~\ref{fig:fgassfr}. The dashed magenta line in each panel represents the median values of $f_{\rm cold}$ in different $\Delta\log{\rm SFR}$ bins.

When comparing the upper and lower panels within identical AGN feedback models, it is evident that the CGM angular momentum does not notably influence the relationship between the gas reservoir and the SFR in SIMBA, in agreement with the result depicted in Figure~\ref{fig:fgassfr}. There is a subtle trend indicating that QGs with higher spins have a slightly stronger dependency of $f_{\rm cold}$ on $\Delta\log{\rm SFR}$. However, due to the significant scatter, the results for low and high spins are consistent with each other, similarly shown in Figure~\ref{fig:fgassfr}, and this holds even when AGN feedback is disabled. 

In summary, the AGN feedback mechanisms will considerably influence the distribution of the CGM angular momentum, as well as the cold gas in the CGM (as illustrated in Figure~\ref{fig:examples}). Nonetheless, the correlation between $f_{\rm cold}$ and $\Delta\log{\rm SFR}$ is not strongly altered by AGN feedback or CGM angular momentum.

\section{Conclusions and discussions}\label{sec:conclusion}

In this research, we explore the impact of CGM angular momentum on star formation processes within galaxies. Employing advanced hydrodynamical simulations from TNG and SIMBA, we have quantified the effects of CGM angular momentum across various AGN feedback scenarios. The key findings of our study are summarised as follows.

(i) The distributions of galaxies in the $\lambda_{\rm CGM}$--$\Delta\log{\rm SFR}$ plane differ significantly between TNG and SIMBA. The most noteworthy distinction between these simulations is that most QGs in TNG exhibit high CGM angular momentum, whereas in SIMBA the opposite is observed. Both simulations show similar distributions of $\lambda_{\rm CGM}$ for SFGs. Additionally, in TNG, QGs generally possess a large cold gas disc in the CGM, and the central cavity in the gas distribution is due to kinetic AGN feedback. However, QGs in SIMBA have minimal cold gas in the CGM and typically lack prominent cold gas discs, in line with their low CGM spin. 

(ii) The angular momentum of the CGM is influenced by both the angular momentum of the environment and the feedback from inner stellar discs (notably the AGN feedback). The SFGs in both TNG and SIMBA exhibit similar strong correlations between $j_{\rm CGM}$ and $j_{\rm Env}$, with a slope near unity for high $j_{\rm Env}$ values. At lower $j_{\rm Env}$, $j_{\rm CGM}$ is more impacted by inner stellar discs and shows a weaker dependence on $j_{\rm Env}$. On the other hand, TNG and SIMBA's QGs show markedly different distributions in the $j_{\rm CGM}$--$j_{\rm Env}$ space. In TNG, QGs have consistently higher $j_{\rm CGM}$ than SFGs with the same $j_{\rm Env}$, resulting from randomised momentum injection events in the kinetic AGN feedback mode. Meanwhile, QGs in SIMBA have median $j_{\rm CGM}$ values similar to SFGs, but the scatter is significantly larger due to collimated jet-mode feedback.

(iii) While the AGN feedback and CGM angular momentum affect the cold gas distribution within the CGM, the relationship between $f_{\rm cold}$ and $\Delta\log{\rm SFR}$ remains largely consistent across various AGN feedback models in SIMBA50. This indicates that the star formation process steadily relies on the availability of cold gas in the CGM. 

Angular momentum quenching has previously been suggested as another mechanism to quench galaxies \citep{Peng2020}. According to \cite{Lu2022}, the high angular momentum of CGM in the QGs of the TNG might be instrumental in preserving their quenched state. Our findings indicate that the CGM angular momentum distribution in SFGs and QGs is heavily influenced by the AGN feedback mechanisms implemented in various simulations. In SIMBA, the jet-mode AGN feedback primarily depletes the cold gas in the CGM of QGs, resulting in these QGs typically having low CGM spins. This indicates that a high CGM angular momentum is not always essential for QGs. Finally, in Appendix~A, we show that our results are not affected by the simulation resolutions.

Observationally, galaxies that possess relatively large \hi\ discs (i.e. \hi-rich) in comparison to typical galaxies with identical stellar masses have also been found \citep[see e.g.,][]{Lemonias2014,Lutz2018}. These galaxies exhibit high CGM angular momenta that contribute to the stabilisation of cold-gas discs and the reduction in star formation efficiency \citep{Obreschkow2016,Lutz2018}. However, these galaxies remain predominantly SFGs, with high CGM angular momentum supporting the large \hi\ discs, unlike the QGs in TNG. 

The red spiral galaxies appear analogous to the QGs in TNG in reality, characterised by their red central regions and higher \hi\ gas fractions when compared to elliptical galaxies of comparable mass \citep{Hao2019,GuoRui2020,Zhou2021}. Nevertheless, as shown in \cite{WangLan2022}, these red spiral galaxies with higher \hi\ masses often exhibit bluer outer discs, suggesting that optically selected red spirals are not completely quenched. Using the \hi\ observations from ALFALFA \citep{Haynes2018}, \cite{Guo2021} demonstrated that the optical morphology of a galaxy does not influence the relationship between total \hi\ mass and SFR. Therefore, the primary characteristic of the cold gas discs in the CGM of QGs in TNG is largely due to its unique AGN feedback mechanism \citep{Guo2022,Ma2022}. Future studies of CGM gas and more sophisticated simulation models will offer a deeper understanding of the impact of CGM angular momentum on star formation activities.

\begin{acknowledgements}
We thank the reviewer for the constructive report that significantly improves the presentation of this paper. This work is supported by the National SKA Program of China (grant No. 2020SKA0110100), the CAS Project for Young Scientists in Basic Research (No. YSBR-092), the National Natural Science Foundation of China (No. 12433003), and the science research grants from the China Manned Space Project with NOs. CMS-CSST-2021-A02. W.C. is supported by the STFC AGP Grant ST/V000594/1, the Atracci\'{o}n de Talento Contract no. 2020-T1/TIC-19882 was granted by the Comunidad de Madrid in Spain, and the science research grants were from the China Manned Space Project. W.C. thanks the Ministerio de Ciencia e Innovación (Spain) for financial support under Project grant PID2021-122603NB-C21 and HORIZON EUROPE Marie Sklodowska-Curie Actions for supporting the LACEGAL-III project with grant number 101086388. We acknowledge the use of the High Performance Computing Resource in the Core Facility for Advanced Research Computing at the Shanghai Astronomical Observatory.
\end{acknowledgements}

%
  \bibliographystyle{aa} 
  \bibliography{ref} 

\begin{thebibliography}{49}
\expandafter\ifx\csname natexlab\endcsname\relax\def\natexlab#1{#1}\fi

\bibitem[{{Angl{\'e}s-Alc{\'a}zar} {et~al.}(2017){Angl{\'e}s-Alc{\'a}zar}, {Dav{\'e}}, {Faucher-Gigu{\`e}re}, {{\"O}zel}, \& {Hopkins}}]{Angles2017}
{Angl{\'e}s-Alc{\'a}zar}, D., {Dav{\'e}}, R., {Faucher-Gigu{\`e}re}, C.-A., {{\"O}zel}, F., \& {Hopkins}, P.~F. 2017, \mnras, 464, 2840

\bibitem[{{Angl{\'e}s-Alc{\'a}zar} {et~al.}(2015){Angl{\'e}s-Alc{\'a}zar}, {{\"O}zel}, {Dav{\'e}}, {Katz}, {Kollmeier}, \& {Oppenheimer}}]{Angles2015}
{Angl{\'e}s-Alc{\'a}zar}, D., {{\"O}zel}, F., {Dav{\'e}}, R., {et~al.} 2015, \apj, 800, 127

\bibitem[{{Appleby} {et~al.}(2021){Appleby}, {Dav{\'e}}, {Sorini}, {Storey-Fisher}, \& {Smith}}]{Appleby2021}
{Appleby}, S., {Dav{\'e}}, R., {Sorini}, D., {Storey-Fisher}, K., \& {Smith}, B. 2021, \mnras, 507, 2383

\bibitem[{{Bacon} {et~al.}(2001){Bacon}, {Copin}, {Monnet}, {Miller}, {Allington-Smith}, {Bureau}, {Carollo}, {Davies}, {Emsellem}, {Kuntschner}, {Peletier}, {Verolme}, \& {de Zeeuw}}]{Bacon2001}
{Bacon}, R., {Copin}, Y., {Monnet}, G., {et~al.} 2001, \mnras, 326, 23

\bibitem[{{Barnes} \& {Efstathiou}(1987)}]{Barnes1987}
{Barnes}, J. \& {Efstathiou}, G. 1987, \apj, 319, 575

\bibitem[{{Bondi}(1952)}]{Bondi1952}
{Bondi}, H. 1952, in Liege International Astrophysical Colloquia, Vol.~4, Liege International Astrophysical Colloquia, ed. P.~{Swings}, 332--336

\bibitem[{{Bullock} {et~al.}(2001){Bullock}, {Dekel}, {Kolatt}, {Kravtsov}, {Klypin}, {Porciani}, \& {Primack}}]{Bullock2001}
{Bullock}, J.~S., {Dekel}, A., {Kolatt}, T.~S., {et~al.} 2001, \apj, 555, 240

\bibitem[{{Cappellari} {et~al.}(2007){Cappellari}, {Emsellem}, {Bacon}, {Bureau}, {Davies}, {de Zeeuw}, {Falc{\'o}n-Barroso}, {Krajnovi{\'c}}, {Kuntschner}, {McDermid}, {Peletier}, {Sarzi}, {van den Bosch}, \& {van de Ven}}]{Cappellari2007}
{Cappellari}, M., {Emsellem}, E., {Bacon}, R., {et~al.} 2007, \mnras, 379, 418

\bibitem[{{Dav{\'e}} {et~al.}(2019){Dav{\'e}}, {Angl{\'e}s-Alc{\'a}zar}, {Narayanan}, {Li}, {Rafieferantsoa}, \& {Appleby}}]{Dave2019}
{Dav{\'e}}, R., {Angl{\'e}s-Alc{\'a}zar}, D., {Narayanan}, D., {et~al.} 2019, \mnras, 486, 2827

\bibitem[{Driver {et~al.}(2007)Driver, Allen, Liske, \& Graham}]{Driver2007}
Driver, S.~P., Allen, P.~D., Liske, J., \& Graham, A.~W. 2007, The Astrophysical Journal, 657, L85

\bibitem[{{Emsellem} {et~al.}(2007){Emsellem}, {Cappellari}, {Krajnovi{\'c}}, {van de Ven}, {Bacon}, {Bureau}, {Davies}, {de Zeeuw}, {Falc{\'o}n-Barroso}, {Kuntschner}, {McDermid}, {Peletier}, \& {Sarzi}}]{Emsellem2007}
{Emsellem}, E., {Cappellari}, M., {Krajnovi{\'c}}, D., {et~al.} 2007, \mnras, 379, 401

\bibitem[{{Fall}(1983)}]{Fall1983}
{Fall}, S.~M. 1983, in Internal Kinematics and Dynamics of Galaxies, ed. E.~{Athanassoula}, Vol. 100, 391--398

\bibitem[{{Fall} \& {Efstathiou}(1980)}]{Fall1980}
{Fall}, S.~M. \& {Efstathiou}, G. 1980, \mnras, 193, 189

\bibitem[{{Fall} \& {Romanowsky}(2013)}]{Fall2013}
{Fall}, S.~M. \& {Romanowsky}, A.~J. 2013, \apjl, 769, L26

\bibitem[{{Guo} {et~al.}(2022){Guo}, {Jones}, \& {Wang}}]{Guo2022}
{Guo}, H., {Jones}, M.~G., \& {Wang}, J. 2022, \apjl, 933, L12

\bibitem[{{Guo} {et~al.}(2021){Guo}, {Jones}, {Wang}, \& {Lin}}]{Guo2021}
{Guo}, H., {Jones}, M.~G., {Wang}, J., \& {Lin}, L. 2021, \apj, 918, 53

\bibitem[{{Guo} {et~al.}(2020){Guo}, {Hao}, {Xia}, {Shi}, {Chen}, {Li}, \& {Gu}}]{GuoRui2020}
{Guo}, R., {Hao}, C.-N., {Xia}, X., {et~al.} 2020, \apj, 897, 162

\bibitem[{{Hao} {et~al.}(2019){Hao}, {Shi}, {Chen}, {Xia}, {Gu}, {Guo}, {Yu}, \& {Li}}]{Hao2019}
{Hao}, C.-N., {Shi}, Y., {Chen}, Y., {et~al.} 2019, \apjl, 883, L36

\bibitem[{{Haynes} {et~al.}(2018){Haynes}, {Giovanelli}, {Kent}, {Adams}, {Balonek}, {Craig}, {Fertig}, {Finn}, {Giovanardi}, {Hallenbeck}, {Hess}, {Hoffman}, {Huang}, {Jones}, {Koopmann}, {Kornreich}, {Leisman}, {Miller}, {Moorman}, {O'Connor}, {O'Donoghue}, {Papastergis}, {Troischt}, {Stark}, \& {Xiao}}]{Haynes2018}
{Haynes}, M.~P., {Giovanelli}, R., {Kent}, B.~R., {et~al.} 2018, \apj, 861, 49

\bibitem[{{Hopkins}(2015)}]{Hopkins2015}
{Hopkins}, P.~F. 2015, \mnras, 450, 53

\bibitem[{{Hopkins} \& {Quataert}(2011)}]{Hopkins2011}
{Hopkins}, P.~F. \& {Quataert}, E. 2011, \mnras, 415, 1027

\bibitem[{{Kennicutt}(1998)}]{Kennicutt1998}
{Kennicutt}, Robert~C., J. 1998, \araa, 36, 189

\bibitem[{{Kere{\v{s}}} {et~al.}(2005){Kere{\v{s}}}, {Katz}, {Weinberg}, \& {Dav{\'e}}}]{Keres2005}
{Kere{\v{s}}}, D., {Katz}, N., {Weinberg}, D.~H., \& {Dav{\'e}}, R. 2005, \mnras, 363, 2

\bibitem[{{Krumholz} \& {Gnedin}(2011)}]{Krumholz2011}
{Krumholz}, M.~R. \& {Gnedin}, N.~Y. 2011, \apj, 729, 36

\bibitem[{{Lemonias} {et~al.}(2014){Lemonias}, {Schiminovich}, {Catinella}, {Heckman}, \& {Moran}}]{Lemonias2014}
{Lemonias}, J.~J., {Schiminovich}, D., {Catinella}, B., {Heckman}, T.~M., \& {Moran}, S.~M. 2014, \apj, 790, 27

\bibitem[{{Lu} {et~al.}(2022){Lu}, {Xu}, {Wang}, {Cai}, {He}, {Xu}, {Xia}, {Mao}, {Springel}, \& {Hernquist}}]{Lu2022}
{Lu}, S., {Xu}, D., {Wang}, S., {et~al.} 2022, \mnras, 509, 2707

\bibitem[{{Lutz} {et~al.}(2018){Lutz}, {Kilborn}, {Koribalski}, {Catinella}, {J{\'o}zsa}, {Wong}, {Stevens}, {Obreschkow}, \& {D{\'e}nes}}]{Lutz2018}
{Lutz}, K.~A., {Kilborn}, V.~A., {Koribalski}, B.~S., {et~al.} 2018, \mnras, 476, 3744

\bibitem[{{Ma} {et~al.}(2022){Ma}, {Liu}, {Guo}, {Cui}, {Jones}, {Wang}, {Zhang}, \& {Dav{\'e}}}]{Ma2022}
{Ma}, W., {Liu}, K., {Guo}, H., {et~al.} 2022, \apj, 941, 205

\bibitem[{{Marinacci} {et~al.}(2018){Marinacci}, {Vogelsberger}, {Pakmor}, {Torrey}, {Springel}, {Hernquist}, {Nelson}, {Weinberger}, {Pillepich}, {Naiman}, \& {Genel}}]{Marinacci2018}
{Marinacci}, F., {Vogelsberger}, M., {Pakmor}, R., {et~al.} 2018, \mnras, 480, 5113

\bibitem[{Mo {et~al.}(1998)Mo, Mao, \& White}]{Mo1998}
Mo, H.~J., Mao, S., \& White, S. D.~M. 1998, Monthly Notices of the Royal Astronomical Society, 295, 319

\bibitem[{{Naiman} {et~al.}(2018){Naiman}, {Pillepich}, {Springel}, {Ramirez-Ruiz}, {Torrey}, {Vogelsberger}, {Pakmor}, {Nelson}, {Marinacci}, {Hernquist}, {Weinberger}, \& {Genel}}]{Naiman2018}
{Naiman}, J.~P., {Pillepich}, A., {Springel}, V., {et~al.} 2018, \mnras, 477, 1206

\bibitem[{{Nelson} {et~al.}(2019{\natexlab{a}}){Nelson}, {Pillepich}, {Springel}, {Pakmor}, {Weinberger}, {Genel}, {Torrey}, {Vogelsberger}, {Marinacci}, \& {Hernquist}}]{dylan2019}
{Nelson}, D., {Pillepich}, A., {Springel}, V., {et~al.} 2019{\natexlab{a}}, \mnras, 490, 3234

\bibitem[{{Nelson} {et~al.}(2018){Nelson}, {Pillepich}, {Springel}, {Weinberger}, {Hernquist}, {Pakmor}, {Genel}, {Torrey}, {Vogelsberger}, {Kauffmann}, {Marinacci}, \& {Naiman}}]{Nelson2018}
{Nelson}, D., {Pillepich}, A., {Springel}, V., {et~al.} 2018, \mnras, 475, 624

\bibitem[{{Nelson} {et~al.}(2019{\natexlab{b}}){Nelson}, {Springel}, {Pillepich}, {Rodriguez-Gomez}, {Torrey}, {Genel}, {Vogelsberger}, {Pakmor}, {Marinacci}, {Weinberger}, {Kelley}, {Lovell}, {Diemer}, \& {Hernquist}}]{Nelson2019}
{Nelson}, D., {Springel}, V., {Pillepich}, A., {et~al.} 2019{\natexlab{b}}, Computational Astrophysics and Cosmology, 6, 2

\bibitem[{{Obreschkow} {et~al.}(2016){Obreschkow}, {Glazebrook}, {Kilborn}, \& {Lutz}}]{Obreschkow2016}
{Obreschkow}, D., {Glazebrook}, K., {Kilborn}, V., \& {Lutz}, K. 2016, \apjl, 824, L26

\bibitem[{{Peebles}(1969)}]{Peebles1969}
{Peebles}, P.~J.~E. 1969, \apj, 155, 393

\bibitem[{{Peng} \& {Renzini}(2020)}]{Peng2020}
{Peng}, Y.-j. \& {Renzini}, A. 2020, \mnras, 491, L51

\bibitem[{{Pillepich} {et~al.}(2018){Pillepich}, {Springel}, {Nelson}, {Genel}, {Naiman}, {Pakmor}, {Hernquist}, {Torrey}, {Vogelsberger}, {Weinberger}, \& {Marinacci}}]{Pillepich2018}
{Pillepich}, A., {Springel}, V., {Nelson}, D., {et~al.} 2018, \mnras, 473, 4077

\bibitem[{{Rahmati} {et~al.}(2013){Rahmati}, {Pawlik}, {Rai{\v{c}}evi{\'c}}, \& {Schaye}}]{Rahmati2013}
{Rahmati}, A., {Pawlik}, A.~H., {Rai{\v{c}}evi{\'c}}, M., \& {Schaye}, J. 2013, \mnras, 430, 2427

\bibitem[{{Romanowsky} \& {Fall}(2012)}]{Romanowsky2012}
{Romanowsky}, A.~J. \& {Fall}, S.~M. 2012, \apjs, 203, 17

\bibitem[{{Schmidt}(1959)}]{Schmidt1959}
{Schmidt}, M. 1959, \apj, 129, 243

\bibitem[{{Springel}(2010)}]{Springel2010}
{Springel}, V. 2010, \mnras, 401, 791

\bibitem[{{Springel} {et~al.}(2018){Springel}, {Pakmor}, {Pillepich}, {Weinberger}, {Nelson}, {Hernquist}, {Vogelsberger}, {Genel}, {Torrey}, {Marinacci}, \& {Naiman}}]{Springel2018}
{Springel}, V., {Pakmor}, R., {Pillepich}, A., {et~al.} 2018, \mnras, 475, 676

\bibitem[{{van de Sande} {et~al.}(2018){van de Sande}, {Scott}, {Bland-Hawthorn}, {Brough}, {Bryant}, {Colless}, {Cortese}, {Croom}, {d'Eugenio}, {Foster}, {Goodwin}, {Konstantopoulos}, {Lawrence}, {McDermid}, {Medling}, {Owers}, {Richards}, \& {Sharp}}]{van2018}
{van de Sande}, J., {Scott}, N., {Bland-Hawthorn}, J., {et~al.} 2018, Nature Astronomy, 2, 483

\bibitem[{{Vogelsberger} {et~al.}(2014){Vogelsberger}, {Genel}, {Springel}, {Torrey}, {Sijacki}, {Xu}, {Snyder}, {Nelson}, \& {Hernquist}}]{Vogelsberger2014}
{Vogelsberger}, M., {Genel}, S., {Springel}, V., {et~al.} 2014, \mnras, 444, 1518

\bibitem[{{Wang} {et~al.}(2022{\natexlab{a}}){Wang}, {Zheng}, {Hao}, {Guo}, {Li}, {Qian}, {Xie}, {Shi}, {Zou}, {Cao}, {Chen}, \& {Xia}}]{WangLan2022}
{Wang}, L., {Zheng}, Z., {Hao}, C.-N., {et~al.} 2022{\natexlab{a}}, \mnras, 516, 2337

\bibitem[{{Wang} {et~al.}(2022{\natexlab{b}}){Wang}, {Xu}, {Lu}, {Cai}, {Xiang}, {Mao}, {Springel}, \& {Hernquist}}]{Wang2022}
{Wang}, S., {Xu}, D., {Lu}, S., {et~al.} 2022{\natexlab{b}}, \mnras, 509, 3148

\bibitem[{{Weinberger} {et~al.}(2017){Weinberger}, {Springel}, {Hernquist}, {Pillepich}, {Marinacci}, {Pakmor}, {Nelson}, {Genel}, {Vogelsberger}, {Naiman}, \& {Torrey}}]{Weinberger2017}
{Weinberger}, R., {Springel}, V., {Hernquist}, L., {et~al.} 2017, \mnras, 465, 3291

\bibitem[{{Zhou} {et~al.}(2021){Zhou}, {Li}, {Hao}, {Guo}, {Mo}, \& {Xia}}]{Zhou2021}
{Zhou}, S., {Li}, C., {Hao}, C.-N., {et~al.} 2021, \apj, 916, 38

\end{thebibliography}
%

\begin{appendix} 
\onecolumn
\section{Resolution effect}
\begin{figure*}[h]
    \centering
    \includegraphics[width=\textwidth]{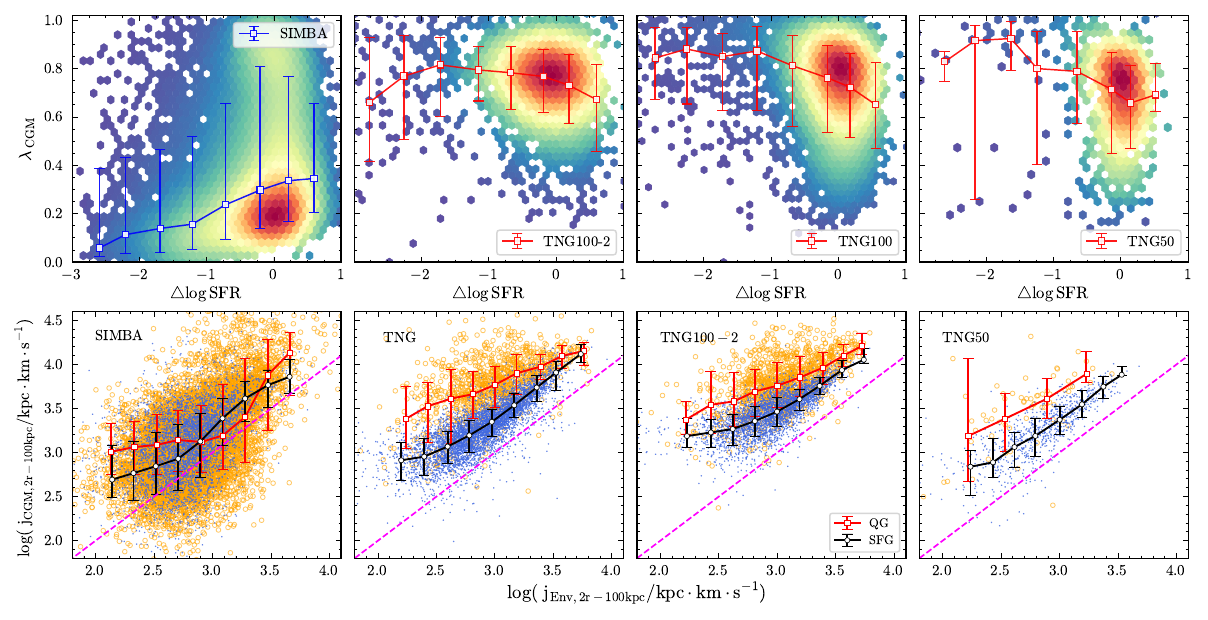}    

    \caption{Same to Figs.~\ref{fig:spinsfr} and ~\ref{fig:jenv}, but also including the results for TNG100-2 and TNG50. Overall, the trends in TNG100-2 and TNG50 are very similar to that in TNG100, with weak dependence on resolution. The quenched galaxies in TNG100-2 and TNG50 still show a higher $\lambda_{\rm CGM}$ than their star-forming counterparts.}
    \label{fig:am_appendix}
\end{figure*} 

In order to check whether our main results will be affected by the resolutions of the specific TNG and SIMBA simulations used in this paper, we investigate the two primary results in TNG100-2 and TNG50. We show the distribution of $\lambda_{\rm CGM}$ as a function of $\Delta\log{\rm SFR}$ and the relations between $\rm j_{CGM}$ and $\rm j_{Env}$, as illustrated in Figure~\ref{fig:am_appendix}. We repeat the results of TNG100 and SIMBA for better comparisons. TNG100-2 has the same box size as TNG100, with mass resolutions lower than TNG100, but very similar to those of SIMBA. The dark matter mass resolution is $m_{\rm DM}=6.0 \times 10^{7}\msun$, and the average gas mass is around $1.1\times 10^{7}\msun$ in TNG100-2. TNG50 simulation has a box size of $35\mpchi$ on a side, with a dark matter mass resolution of $m_{\rm DM}=4.5 \times 10^{5}\msun$ and an average gas mass around $8.5\times 10^{4}\msun$, which are significantly higher than those of TNG100 and TNG100-2. Here, we confirm that, even though the exact values of $\lambda_{\rm CGM}$ and $\rm j_{CGM}$ will be affected by the simulation resolution, our results remain unchanged in simulations of different resolutions.
\end{appendix}

\end{document}